\documentclass[a4paper,twocolumn]{article}
\usepackage[dvips]{graphicx}
\begin{document}

\title{Effects of noisy entangled state transmission\\ 
on quantum teleportation}

\author{Holger F. Hofmann$^{1,2}$, Kenji Tsujino$^2$, 
and Shigeki Takeuchi$^{1,2}$ \\ $^1$ CREST, Japan Science and
Technology Corporation (JST)\\ $^2$ Research Institute for Electronic Science,\\ Hokkaido University, Sapporo 060-0812\\
Tel/Fax: 011-706-2648\\ 
e-mail: h.hofmann@osa.org}

\date{}

\maketitle

\begin{abstract}
Quantum teleportation requires the transmission of entangled
pairs to Alice and Bob. Transmission errors modify the entangled
state before the teleportation can be performed. We determine
the changes in the output state caused by such transmission
errors. It is shown that the errors caused by entanglement
transmission are equivalent to the errors caused in a direct
transmission of a quantum state from Alice to Bob.
\\[0.2cm]

\end{abstract}

\section{Introduction}
Quantum teleportation is an entanglement assisted method for
transfering quantum information \cite{Ben93}. Since  
teleportation strategies have been realized experimentally
\cite{Bou97,Fur98}, 
it is now of interest to discuss possible applications
and their technical limitations. If quantum teleportation is
actually applied to transfer quantum information over large
distances, one of the major sources of errors will be the 
quantum communication lines used for entanglement distribution.
In the following, we present a general formalism by which the 
effect of a specific type of error in the entanglement distribution
on the output state of quantum teleportation can be predicted.
In section \ref{sec:ideal}, we present a formulation of ideal
quantum teleportation for any N-level system. In section 
\ref{sec:transprop}, the general error transfer properties are
analyzed. Finally, the result is analyzed in section
\ref{sec:discuss} and conclusions are 
drawn in section \ref{sec:conclusions}.

\section{Ideal quantum teleportation}
\label{sec:ideal}
Quantum teleportation can be applied to any N-level quantum 
system. Ideally, it requires a maximally entangled state.
Given an appropriate basis $\mid n \rangle$, such a quantum
state can always be written as
\begin{equation}
\mid E_{\mbox{max}} \rangle_{i,j} = 
\frac{1}{\sqrt{N}} \sum_{n=0}^{N-1}
\mid n \rangle_i \otimes \mid n \rangle_j.
\end{equation}
The second requirement for quantum teleportation is the
ability to carry out a measurement projecting a pair of
systems onto a maximally entangled state. Ideally, each 
measurement result $m$ corresponds to a maximally entangled 
state,
\begin{equation}
\mid P(m) \rangle_{i,j} = \sqrt{\frac{\chi(m)}{N}} \sum_{n=0}^{N-1}
\left( \hat{U}(m) \mid n \rangle \right)_i 
\otimes \mid n \rangle_j,
\end{equation}
where the normalization factor $\chi(m)$ and the unitary 
transformation $\hat{U}(m)$ fulfill the conditions for a
positive operator valued measure,
\begin{equation}
\sum_m \mid P(m) \rangle \langle P(m) \mid = \hat{1}.
\end{equation}
Note that this requirement is only necessary for unconditional
teleportation. In conditional teleportation, only some measurement
results $m$ correspond to maximally entangled states. The following
analysis applies to both conditional and unconditional 
teleportation. However, the loss of quantum information due to a
conditional teleportation may easily be a more serious source of
error than the transmission errors considered here.

For a specific measurement result $m$, the principle of quantum 
teleportation may now be represented schematically by assigning
the input state to system $A$ and the entanglement to a reference
system $R$ and the remote system $B$. The measurement is performed
on the joint input from system $A$ and the reference system $R$.
As a result of the entanglement, $B$ is then projected into a
unitary transformation of the input state 
$\mid \psi_{\mbox{in}} \rangle$
conditioned by the measurement result $m$, as given by
\begin{equation}
\label{eq:grid}
\begin{array}{cccccc}
\multicolumn{6}{l}{\mbox{Initial state}}\\[0.2cm]
\frac{1}{\sqrt{N}}\sum_n & \mid \psi_{\mbox{in}} \rangle_A 
& \otimes 
& \mid n \rangle_R & \otimes & \mid n \rangle_B \\[0.3cm]
\multicolumn{6}{l}{\mbox{Measurement projection}}\\[0.2cm]
\sqrt{\frac{\chi(m)}{N}} \sum_n & \hspace{-0.4cm}
\left(\langle n \mid \hat{U}^{-1}(m)\right)_A \hspace{-0.3cm}
& \otimes & \langle n \mid_R& & \\[0.3cm]
\multicolumn{6}{l}{\mbox{Conditional output state}}\\[0.2cm]
\frac{\sqrt{\chi(m)}}{N} \sum_n  & & 
\multicolumn{3}{c}{\hspace{-3cm}\langle n \mid 
\hat{U}^{-1}(m)\mid \psi_{\mbox{in}}\rangle}
& \mid n \rangle_B.\\[0.5cm]
\multicolumn{6}{c}{\hspace{-0.5cm} = \frac{\sqrt{\chi(m)}}{N}
\hat{U}^{-1}(m)\mid \psi_{\mbox{in}}\rangle_B}
\end{array}
\end{equation}
It should be noted that, at this point, no information has been
transfered from $A$ to $B$ and the density matrix of an observer
in $B$ is still in a maximally mixed state as long as the
measurement result $m$ is not known. Teleportation is completed
by classically communicating the measurement result $m$ to $B$.
The unitary transformation $\hat{U}^{-1}(m)$ can then be 
compensated by performing $\hat{U}(m)$ on the output to restore
the original state $\mid \psi_{\mbox{in}}\rangle$ in $B$.

\section{Error transfer}
\label{sec:transprop}
\begin{figure}
\hspace{0.3cm}
\begin{picture}(220,240)
\put(10,205){\makebox(60,20){
          $\mid \psi_{\mbox{in}}\rangle_{\mbox{A}}$}}
\put(40,200){\line(0,-1){30}}
\put(40,170){\line(1,2){7}}
\put(40,170){\line(-1,2){7}}
\put(5,132){\framebox(70,36){}}
\put(5,150){\makebox(70,18){Measurement}}
\put(5,132){\makebox(70,18){Output $=m$}}

\put(40,130){\line(1,-2){10}}
\put(70,70){\line(-1,2){12}}
\put(40,130){\line(0,-1){12}}
\put(40,130){\line(3,-2){10}}

\put(39,92){\makebox(30,20){$\hat{F}_R$}}

\put(180,130){\line(-1,-2){12}}
\put(150,70){\line(1,2){13}}
\put(180,130){\line(0,-1){12}}
\put(180,130){\line(-3,-2){10}}

\put(151,92){\makebox(30,20){$\hat{F}_B$}}

\put(70,34){\framebox(80,36){}}
\put(70,52){\makebox(80,18){Source of}}
\put(70,34){\makebox(80,18){Entanglement}}

\put(75,148){\line(1,0){80}}
\put(75,152){\line(1,0){80}}
\put(159,150){\line(-2,1){10}}
\put(159,150){\line(-2,-1){10}}

\put(85,165){\makebox(60,8){\small classical}}
\put(85,157){\makebox(60,8){\small communication}}

\put(150,130){\makebox(60,36){$\hat{U}(m)$}}
\put(180,148){\circle{30}}

\put(180,200){\line(0,-1){30}}
\put(180,200){\line(1,-2){7}}
\put(180,200){\line(-1,-2){7}}

\put(130,205){\makebox(90,20){
   $\hat{F}_{\mbox{out}}(m) 
          \mid \psi_{\mbox{in}}\rangle_{\mbox{B}}$}}
\end{picture}
\vspace{-1cm}
\caption{\label{schematic} Schematic representations of
entanglement distribution errors in quantum teleportation.}
\end{figure}
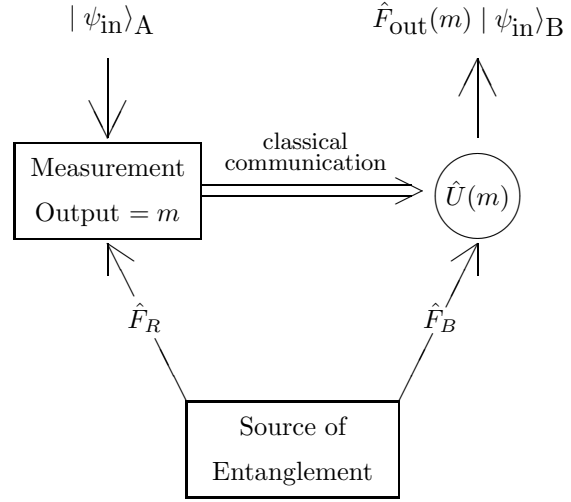

For long distance teleportation, the most important technical 
requirement is the distribution of entanglement over long 
distances. Transmission errors act randomly on the transmission
lines for systems $R$ and $B$. However, all such random errors
can be represented as mixtures of quantum mechanically precise
operators $\hat{F}_R(x_r)$ and $\hat{F}_B(x_b)$, where $x_r$ and
$x_b$ are the random variables defining the type of error.
The probability distribution over $x_r$ and $x_b$ is given by
the expectation values of 
$\hat{F}^\dagger_i(x_i)\hat{F}_i(x_i)$, requiring that
\begin{equation}
\sum_{x_i} \hat{F}^\dagger_i(x_i)\hat{F}_i(x_i) = \hat{1}.
\end{equation}
Figure \ref{schematic} illustrates the effect of a well defined
pair of errors $\hat{F}_R(x_r)$ and $\hat{F}_B(x_b)$ on the 
teleportation process. The total error can be represented by an
output error operator $\hat{F}_{\mbox{out}}(m;x_r,x_b)$.
This operator can be derived from the teleportation scheme
introduced in section \ref{sec:ideal}.

With the errors, the initial state for teleportation 
reads
\begin{equation}
\frac{1}{\sqrt{N}}\sum_n \mid \psi_{\mbox{in}} \rangle_A 
\otimes 
\hat{F}_R(x_r)\mid n \rangle_R \otimes 
\hat{F}_B(x_b)\mid n \rangle_B.
\end{equation}
By replacing the ideal initial state in equation (\ref{eq:grid}),
the outcome of the teleportation measurement of $m$ can be
calculated. It reads
\begin{eqnarray}
\lefteqn{\frac{\sqrt{\chi(m)}}{N} \sum_{n,n^\prime}
\langle n^\prime\mid \hat{U}^{-1}(m)\mid \psi_{\mbox{in}}\rangle}
\nonumber \\ && \times
\langle n^\prime \mid \hat{F}_R(x_r)\mid n \rangle \hat{F}_B(x_b)
\mid n \rangle_B.
\end{eqnarray}
This equation may be greatly simplified by intruducing the
transpose $\hat{F}_R^T(x_r)$ of $\hat{F}_R(x_r)$ in the 
$\mid n \rangle$ basis, defined by
\begin{equation}
\langle n^\prime \mid \hat{F}_R(x_r)\mid n \rangle
= \langle n \mid \hat{F}_R^T(x_r)\mid n^\prime \rangle.
\end{equation}
With this definition, the output state of the teleportation
after the application of $\hat{U}(m)$ 
reads
\begin{equation}
\frac{\sqrt{\chi(m)}}{N} \hat{U}(m)
\hat{F}_B(x_b)\hat{F}_R^T(x_r)
\hat{U}^{-1}(m)\mid \psi_{\mbox{in}}\rangle_B.
\end{equation}
The difference between this teleportation result and the
ideal result is described by the effective error operator
$\hat{F}_{\mbox{out}}(m;x_r,x_b)$, given by
\begin{equation}
\label{eq:eff}
\hat{F}_{\mbox{out}}(m;x_r,x_b) = \hat{U}(m)
\hat{F}_B(x_b)\hat{F}_R^T(x_r)
\hat{U}^{-1}(m).
\end{equation}
With this equation, it is possible to characterize the
teleportation errors based on any combination of 
transmission errors $\hat{F}_R(x_r)$ and  $\hat{F}_B(x_b)$.
In general, the errors are only changed by unitary transformations,
indicating that any properties invariant under such transformations
will be preserved. All other properties of an error are
transformed according to the measurement result $m$ obtained
in the teleportation process. Knowledge of $m$ is therefore 
important in order to minimize the effects of errors in 
the teleportation process \cite{Hof00}.

\section{Properties of the output error}
\label{sec:discuss}

The result shows that the errors in $R$ and the errors in
$B$ are effectively multiplied, representing the expected
exponential increase in errors with the length of the 
communication line. Indeed, the total error may be interpreted
as a sequential error of an initial transmission through
channel $R$ with an error of 
$\hat{U}(m)\hat{F}^T_R(x_r)\hat{U}^{-1}(m)$, followed
by a transmission through channel $B$ with an error of
$\hat{U}(m)\hat{F}_B(x_b)\hat{U}^{-1}(m)$. The only differences
between the teleportation and the direct transmission of 
the quantum state through the transmission lines used for
$R$ and $B$ are given by the unitary transformation 
$\hat{U}(m)$ and the transposition operation on the error in
$R$. However, this transposition is also a unitary transformation
that preserves the general properties of $\hat{F}_R(x_r)$, such
as the eigenvalues or the overlap between eigenstates. 
From this property, it is clear that the channel capacity of
quantum teleportation is exactly identical with the capacity
for a direct transmission. It is for this reason, that the use of
teleportation as a quantum repeater requires some form of
reliable entanglement purification \cite{Dur99}. Nevertheless
it is interesting to note that the errors are not increased by
the use of entanglement in the channels. In this context,
entanglement is no more sensitive to errors then any arbitrary
quantum state.  

It is now possible to consider different types of error
statistics. The most simple errors are homogeneous ones,
such that any unitary transformation $\hat{U}(m)$ merely
causes a permutation of the errors $x_i$, 
\begin{equation}
\hat{U}(m)\hat{F}_i(x_i)\hat{U}^{-1}(m) = \hat{F}_i(x^\prime_i).
\end{equation}
If the $x_i$ are unknown, such errors are independent of $m$
and the teleportation errors correspond to the direct 
transmission except for the transposition in $R$. However,
it is likely that errors homogeneous with respect to
$\hat{U}(m)$ are also homogeneous with respect to the 
transposition,
\begin{equation}
\hat{F}^T_R(x_r) = \hat{F}_R(x^\prime_r).
\end{equation}
In fact, a much wider class of errors should fulfill this 
requirement, making it possible to identify the transposed
errors with non-transposed ones. 

In order to illustrate the physics of a transposition, it
is helpful to consider the angular momentum operators
$\hat{l}_x$,$\hat{l}_y$, and $\hat{l}_z$. A transposition
in the basis of $\hat{l}_z$ eigenstates leaves both 
$\hat{l}_x$ and $\hat{l}_z$ invariant, while reversing 
the sign of $\hat{l}^T_y=-\hat{l}_y$. The transposition
therefore corresponds to a mirror image in the $xz$-plane.
For reasons of symmetry it is likely that transmission
errors will indeed be homogeneous with respect to this
operation. In the case of transmission lines with a 
non-homogeneous error, the transposition will effectively
reverse the asymmetry.

For errors which are not homogeneous with respect to 
$\hat{U}(m)$, the output error statistics depend strongly
on the measurement result $m$. Equation (\ref{eq:eff}) then
describes the correlation between $m$ and the output error.
If the output states are analyzed without reference to $m$,
the teleportation will be far more noisy than a direct 
transmission. However, this type of noise originates from
a classical randomization of the signal based on the 
known variable $m$. This noise source can therefore be 
eliminated by analyzing the full correlated statistics.

\section{Conclusions}
\label{sec:conclusions}
The entanglement distribution errors are essentially equivalent
to errors in a direct transmission of quantum states through
the same communication lines. If the errors are homogeneously
distributed, it makes no difference whether the quantum state
is teleported or it is sent directly. For non-homogeneous
errors, the teleportation measurement result $m$ represents
an additional source of randomness. However, the random variable
$m$ is known to both the sender and the receiver, indicating
that this randomness will not cause any loss of information.
In conclusion, the errors caused by entanglement distribution
are remarkably similar to the errors of a direct transmission
through the same quantum channels. The method of teleportation
itself neither reduces nor enhances the errors.




\begin{thebibliography}{5}
\bibitem{Ben93}
C.H. Bennet, G. Brassard, C. Crepeau, R.Jozsa, A. Peres, and W.K. Wootters, 
Phys. Rev. Lett. {\bf 70}, 1895 (1993).

\bibitem{Bou97}
D. Bouwmeester, J.-W. Pan, K.Mattle, M.Eibl, H. Weinfurter,
and A. Zeilinger, Nature (London) {\bf 390}, 575 (1997).

\bibitem{Fur98}
A. Furusawa, J.L. Sorensen, S.L. Braunstein, C.A. Fuchs, H.J. Kimble, and 
E.S. Polzik, Science {\bf 282}, 706 (1998).

\bibitem{Hof00}
H.F. Hofmann, T. Ide, T. Kobayashi, and A. Furusawa, Phys. Rev. A {\bf 62},
062304 (2000).

\bibitem{Dur99}
W.D\"ur, H.-J. Briegel, J.I. Cirac, and
P. Zoller, Phys. Rev. A {\bf 59}, 169 (1999).

\end{thebibliography}
\end{document}